\documentclass[preprint,12pt,a4paper,byrevtex]{revtex4}%
\usepackage{natbib}
\usepackage{graphicx}
\usepackage{amsfonts}
\usepackage{amsmath}
\usepackage{amssymb}%
\providecommand{\U}[1]{\protect\rule{.1in}{.1in}}

\begin{document}
\preprint{ }
\preprint{ }
\title{Very strong reduction of 1/f noise by Carbon doping in epitaxial
Fe/MgO(100)12ML/Fe magnetic tunnel junctions with large density of barrier defects}
\author{D. Herranz, R.Guerrero, J.P.Cascales, F.G.Aliev}
\affiliation{Dpto F\'{\i}sica Materia Condensada, C-III, Universidad Aut\'{o}noma de
Madrid, Cantoblanco, Spain}
\author{M. Hehn, C. Tiusan}
\affiliation{Institut Jean Lamour, Nancy Universit\'{e}, 54506 Vandoeuvre-l\`{e}s-Nancy
Cedex, France}
\pacs{PACS numbers}

\begin{abstract}
We report on the strong influence of Carbon doping on 1/f noise in
fully epitaxial Fe/MgO(100)12ML/Fe magnetic tunnel junctions in
comparison with undoped junctions with a large density of barrier
defects. Carbon influences the relaxation of defects, the
reconstruction of the interface and the symmetry transformation of
interface resonance states, which are suggested to contribute to the
strong reduction of the 1/f noise. Our study demonstrates that
doping with light elements could be a versatile tool to improve the
electron transport and noise in epitaxial magnetic tunnel junctions
with large density of barrier defects.

\end{abstract}

\maketitle

The realization that tunneling magnetoresistance (TMR) depends on the atomic structure of the entire junction \cite{DeTeresa}, was followed by the discovery of coherent tunneling phenomena in magnetic tunnel junctions (MTJ) based on MgO barriers\cite{Butler2001,Mathon,Parkin,Yuasa2004}. Until now, the main challenge that researchers have faced has been improving the TMR \cite{Ikeda2008} and its dependence on the applied bias \cite{Tiusan2006}. Since it is relevant for applications, the next strategic objective is to minimize the 1/f noise in MTJs, particularly ubiquitous for low dimensional materials and nanoscale devices \cite{Jiang2004,Guerrero2005,Ferreira2006,Almeida2006}. Besides annealing \cite{Stearrett2010
} and e-beam
evaporation \cite{Diao2010}, mechanisms that minimize 1/f noise in these structures remain poorly explored. An effective way to optimize the characteristics of magnetic tunnel junctions is doping them with light elements, such as Carbon \cite{Tiusan2006}, Boron \cite{Lu2009} or Nitrogen \cite{
herranz2010}. Regarding Carbon doping, we observed that for MTJs with 10ML MgO barriers and a very low density of barrier defects, Carbon doping produces about a 50\% increase of the average normalized 1/f noise \cite{Aliev2007}. For thick MgO barriers, 
a variation of barrier thickness from 10 to 12 ML 
increases the density of defects in the barrier, on the roughness of the top MgO interface and on the in-plane structural coherence of the insulator.

In this study, we report 
a strong reduction of 1/f noise with 
Carbon doping 
in Fe/MgO(12ML)/Fe MTJs
, with MgO thicknesses well above the critical value where a plastic relaxation occurs and dislocations nucleate due to the Fe/MgO lattice mismatch \cite{Tiusan2007}.  These
effects have 
consequences on the tunneling and noise. We discuss the relation between symmetry and electron structure modification of the FeC/MgO\ interface in order to explain our main experimental findings.

\begin{center}
\begin{figure}[h]
    \begin{center}
    \includegraphics[width=12cm]{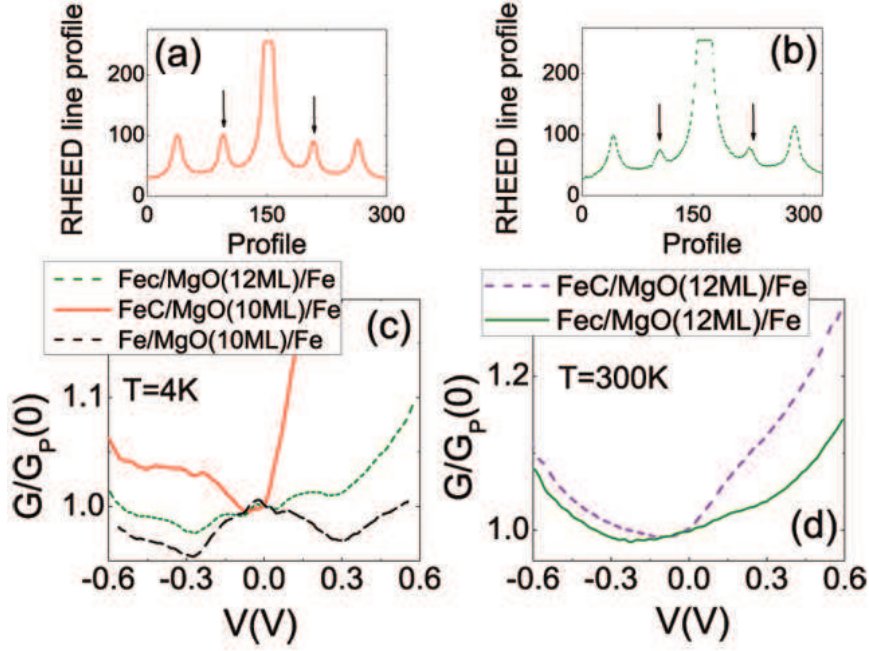}
    \caption{RHEED profiles for Carbon doped MTJs with 10ML (a) and 12 ML (b) thick MgO barriers. Part (c) shows typical parallel dynamic conductances in MTJs with different concentrations of Carbon, measured at T=4K and normalized by the conductance in the P state at 0 V bias. Part (d) compares the bias dependence of the dynamic conductance, at 300K, in the P state for two MTJ2 junctions with 12ML of MgO barriers, marked by dotted and continuous arrows in the \emph{Fig. \ref{fig:f3} (b)}.} \label{fig:f1}
\end{center}
\end{figure}
\end{center}

Two types of 
MTJs were grown by Molecular Beam Epitaxy (MBE) on MgO (100) substrates under ultra high vacuum (UHV) conditions, typically at a base pressure of $5\cdot10^{-11}$ mbar. The first type (MTJ1) has the following structure: MgO$\left(100\right)  $ //MgO$(10nm)$/ Fe$(45nm)$/ MgO$(12ML)$/ Fe$(10nm)$/ Co$(20nm)$/ Pd$(10nm)$/ Au$(10nm)$. For the second type of junctions (MTJ2) the\ 10nm MgO\ buffer layer over the MgO\ substrate is absent. Carbon impurities are already present in the MgO substrate before the growth. The initial annealing does not completely remove the carbon impurities from the substrate. Therefore when the Fe layer is directly grown over the MgO\ substrate
\cite{Tiusan2007}, the Carbon diffuses into the ferromagnetic electrode and MgO barrier and may even segregate to the top Fe surface. Two-dimensional layer-by-layer growth was observed up to several monolayers by means of reflection high-energy electron diffraction (RHEED) intensity oscillations along the [100] direction. The RHEED analysis reveals a c(2$\times$2)-type reconstruction pattern for MTJ2 samples (marked peaks in
\emph{Fig. \ref{fig:f1} (a),(b)}), absent for 
MTJ1 samples\cite{Tiusan2007}. When MgO is grown on Fe(001), after the critical thickness of 5-6 ML a plastic relaxation occurs inducing dislocations within the barrier \cite{
Tiusan2007}. After the relaxation, the number of defects within the barrier and the roughness of the top MgO interfaces increases with the barrier thickness. Phase shift Transmission Electron Microscopy analysis \cite{Tiusan2007} shows that dislocations within the MgO barrier have an oblique orientation. This implies that for a specific
density of defects, increasing the barrier thickness reduces the defect-free junction area where coherent tunneling with symmetry filtering occurs. 
Square shaped MTJs with areas A=$10\cdot10\mu m^{2}$ and $30\cdot30\mu m^{2}$ have been patterned, using a standard optical lithography/ion etching process, controlled step by step \textit{in situ} by Auger spectroscopy. Conductance and low frequency noise were measured using the experimental setup previously described in Ref.\cite{Guerrero2005}

For MTJ1 samples, in agreement with previous observations \cite{Tiusan2007}, the dynamic conductance in the parallel (P) state (G$_{p}$(V)) shows an excess in conductance at low bias (below 0.3V), due to an excess of tunneling current into the $\Delta_{5}$ interface or bulk states. Carbon doping qualitatively changes the G$_{p}$(V) dependence and turns the dynamic conductivity vs bias behavior more asymmetric\cite{Tiusan2007}. The sample in Fig.1(a) shows a higher Carbon content than (b), indicated with C and c in \emph{Fig. \ref{fig:f1} (c)} respectively. Consequently, the conductance for sample (b) is more similar to the Carbon free conductance, while for sample (a) the behavior is more asymmetric (see Fig.1(c)). This asymmetry is most probably induced by the spatial distribution of the Carbon diffusion. The other important effect due to Carbon doping is the suppression of the low bias 
excess 
G(V) (\emph{Fig. \ref{fig:f1} (c)}). \emph{Figure \ref{fig:f1} (d)} represents the G$_{p}$(V) curves measured at 300K for two different MTJ2 samples indicating the possible variation of Carbon concentration within the MTJ set with the same MgO barrier thickness (see arrows in \emph{Fig. \ref{fig:f3} (b)}).

\begin{center}
\begin{figure}[h]
    \begin{center}
    \includegraphics[width=12cm]{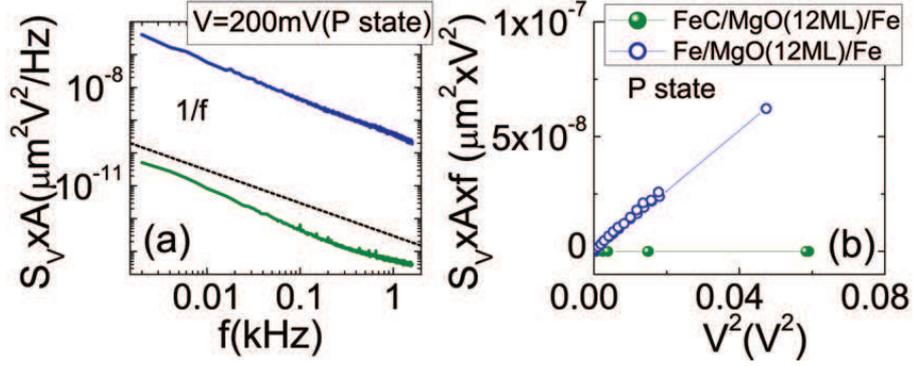}
    \caption{(a) Frequency dependence of the typical normalized noise power for Carbon free (dotted) and doped (full lines) MTJs. Part (b) shows proportionality between the noise power and the bias squared for Carbon doped and undoped MTJs. Black dotted line represents 1/f pure dependence.} \label{fig:f2}
\end{center}
\end{figure}
\end{center}

\begin{center}
\begin{figure}[h]
    \begin{center}
    \includegraphics[width=10cm]{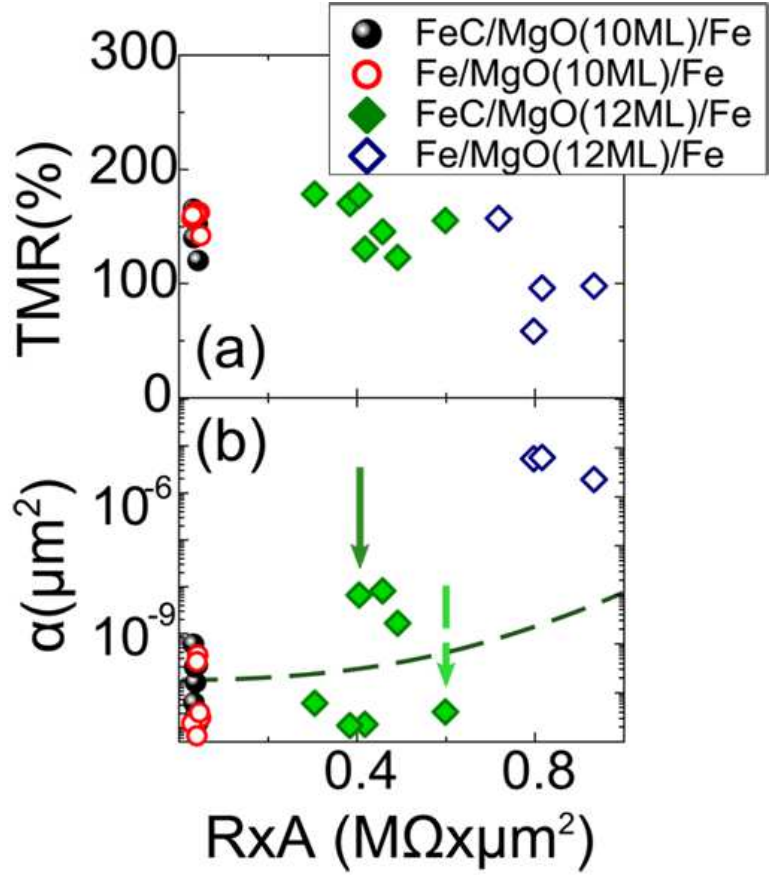}
    \caption{TMR (a) and Hooge factor (b) as a function of R$\times$A product for MTJs(1,2) with 10\cite{Aliev2007} and 12 ML thick MgO barriers. The dotted line represents the mean empirical value of $\alpha$ reported in Ref.\cite{Gokce2006}. The green and purple arrows represent the samples of \emph{Fig. \ref{fig:f1} (d)} respectively.}\label{fig:f3}
\end{center}
\end{figure}
\end{center}

The typical noise power spectra (\emph{S}$_{V}$) 
in the P state (\emph{Fig. \ref{fig:f2} (a)}) reveal the presence of 1/f noise. Indeed, in the frequency range between 1 and 50 Hz, 
we have that $S_{V}(f)\propto 1/f^{\beta}$ (with $0.5< \beta <1.5$) \cite{1fnoise}, allowing us to describe the noise as 1/f-like. The bias dependence of the normalized 1/f noise can be determined through the Hooge factor ($\alpha$) from the phenomenological expression: \emph{S}$_{V}$\emph{(f)=}$\alpha\cdot$%
\emph{(I}$\cdot$\emph{R)}$^{2}$\emph{/(A}$\cdot$\emph{f)}, where \emph{R} is the resistance, \emph{I} is the current, \emph{A} is the area and \emph{f} is the frequency. We also note that $\alpha$\emph{(V)} correlates with TMR(V), indicating that the origin of the 1/f noise is due to symmetry dependent tunneling resistance fluctuations. As can be seen in \emph{Fig. \ref{fig:f2} (b)}, the average low frequency noise power increases approximately as $V^{2}$ 
, which indicates dominant 1/f noise. \emph{Fig. \ref{fig:f2}} shows a strong suppression of 1/f noise with Carbon doping.

\emph{Figure \ref{fig:f3}} summarize the TMR and $\alpha$ in the P state (
at 200mV
), respectively, as a function of the R$\cdot$A product. 
From the one hand, both MTJs1,2 with 10ML MgO barrier show similar values of low frequency noise and TMR \cite{Aliev2007}. 
In contrast, MTJ2 samples with 12ML of MgO show much lower 1/f noise levels than MTJ1 junctions. Besides, the R$\cdot$A for MTJ2 samples is about two times smaller than for MTJ1 while the average TMR is somewhat higher for MTJ2. The dotted line represents empirical values of the average $\alpha$ reported in
\cite{Gokce2006}, clearly indicating that Carbon doping in Fe/MgO(12ML)/Fe MTJs strongly decreases the 1/f noise while increasing or maintaining a high TMR level.


The low sensitivity of the 1/f noise level in Fe/MgO(10ML)/Fe to Carbon could implicate a reduced number of defects even without doping. Then, the dominant tunneling is coherent across single crystal areas of the MTJ between dislocations, with conservation of k$_{\parallel}$ and symmetry filtering effects. The situation seems to be very different in 
thick MTJs
. For 12ML MgO thickness, interface Carbon doping substantially decreases the 1/f noise and increases the G$_{p}$(V) (in about a factor of 2). This indicates that additional conductivity channels open, most probably related to dislocations/defects and tunneling where k$_{\parallel}$ conservation is violated.

Significant differences in noise behavior and the response to carbon doping for systems with barriers varying from 10 to 12 ML could be explained as follows. The thickness range of 10 and 12 MgO ML is well above the plastic relaxation when growing the MgO on Fe. In this regime, the variation of barrier thickness 
has a drastic effect both on the density of defects in the barrier, on the roughness of the top MgO interface and on the in-plane structural coherence of the insulator. The reduction of the lateral structural coherence of the insulator has been experimentally certified by the damping of the RHEED oscillations when monitoring the growth of MgO on Fe
\cite{herranz2010}. 
The reduction of the in-plane coherence by increasing the density of defects will have drastic effects on the tunneling and
tunneling fluctuations (noise). Therefore, since 12 ML thick MgO barriers show a significantly higher density of defects than 10 ML barriers, the effect of C doping is expected to be larger.

Recent band structure calculations \cite{Lu2011} suggest the change of symmetry of interface resonant states (IRS) from $\Delta_{5}$ to $\Delta_{1}$ with Carbon doping. Experimentally, this change is reflected in the suppression of the excess zero bias conductance (\emph{Fig. \ref{fig:f1} (c)}). If we assume that the 1/f noise in the P state is originated from temporal fluctuations of the interface atoms, with uniformly distributed relaxation times, which scatter dominating $\Delta_{1}$ electrons into the $\Delta_{5IRS}$ states or vice versa. Then, for MTJ1, we could roughly estimate the related normalized fluctuations of conductance as follows: $\langle[\sigma(\Delta_{1})-\sigma(\Delta_{5IRS})]^{2}\rangle/\langle\sigma^{2}(\Delta_{1})\rangle \sim1$ (with $\sigma(\Delta_{5IRS})<<\sigma(\Delta_{1})$). At the same time, for the 
MTJ2 with IRS dominated by $\Delta_{1}$ symmetry, fluctuations in the position of interface atoms provide much smaller fluctuations of the conductance:
$\langle[\sigma(\Delta_{1})-\sigma(\Delta_{1IRS})]^{2}\rangle/\langle[\sigma(\Delta_{1})+\sigma(\Delta_{1IRS})]^{2}\rangle \ll1 $ (in the condition $\sigma(\Delta_{1IRS})\simeq\sigma(\Delta_{1})$).

Some other observations, such as the robustness of the doped MTJs at high voltages \cite{Guerrero2007}, indicate that other mechanisms could also contribute to the strong suppression of the 1/f noise
. One possible scenario is that the Carbon suppresses Fe-O interdiffusion \cite{wang2011} and relaxes the MgO barrier. Besides, fluctuating trap charges in the extended defects inside MgO could be partially \textquotedblleft filled\textquotedblright\ and relaxed by the Carbon doping, strongly reducing the 1/f noise.

The spatial distribution of the Carbon concentration parallel to the interfaces (experimentally confirmed by in-situ RHEED analysis of the bottom Fe(001) surface) could induce a local variation of the Fe/MgO interface reconstruction between different MTJs from the same substrate. By analyzing
differences in the low bias $G_{p}(V)$ behavior between FeC/MgO(12ML)/Fe MTJs belonging to two groups with distinct 1/f noise levels (Figure 1(d)), and based on the idea of the suppression of the $\Delta_{5}$ IRS by Carbon doping \cite{Lu2011}, we identify the presence of a higher Carbon concentration in MTJs with the lowest 1/f noise values (\emph{Fig. \ref{fig:f3} (b)}).

In conclusion, Carbon doping of epitaxial Fe/MgO/Fe MTJs with MgO barriers well above the critical thickness for plastic relaxations
, substantially decreases the 1/f noise and improves MTJ robustness. Our study demonstrates that doping with light elements could be a new promising tool to gain control over electron transport and noise in epitaxial magnetic tunnel junctions.

The work was supported by Spanish MICINN (MAT2009-10139, CSD2007-00010, FR2009-0010) and CAM (P2009/MAT-1726).

\end{document}